\documentstyle[12pt,epsf,epsfig,amsmath]{article}
\newfont{\thiplo}{msbm10 scaled\magstep 2}
\newfont{\gothic}{eufb10 scaled\magstep 2}
\newfont{\unc}{eurb10} 
\newskip\humongous \humongous=0pt plus 1000pt minus 1000pt
\def\caja{\mathsurround=0pt}\def\eqalign#1{\,\vcenter{\openup1\jot \caja
        \ialign{\strut \hfil$\displaystyle{##}$&$ 
        \displaystyle{{}##}$\hfil\crcr#1\crcr}}\,}
\newif\ifdtup

\def\eqright #1\cr{\noalign{\hfill$\displaystyle{{}#1}$}}
\def\eqleft #1\cr{\noalign{\noindent$\displaystyle{{}#1}$\hfill}}

\def\oldreffmt#1{\rlap{[#1]} \hbox to 2\parindent{}}

\def\figfmt#1{\rlap{Figure {#1}} \hbox to 1in{}}

\def\sectioneq{\def\theequation{\thesection.\arabic{equation}}{\let
\holdsection=\section\def\section{\setcounter{equation}{0}\holdsection}}}%

\newcounter{holdequation}

\def\auto{\eqno(\refstepcounter{equation}\theequation)}
\def\begineq #1\endeq{$$ \refstepcounter{equation}\eqalign{#1}\eqno
	(\theequation) $$}
\def\contlimit{\,{\hbox{$\longrightarrow$}\kern-1.8em\lower1ex
\hbox{${\scriptstyle (a\rightarrow0)}$}}\,}
\def\centeron#1#2{{\setbox0=\hbox{#1}\setbox1=\hbox{#2}\ifdim
\wd1>\wd0\kern.5\wd1\kern-.5\wd0\fi
\copy0\kern-.5\wd0\kern-.5\wd1\copy1\ifdim\wd0>\wd1
\kern.5\wd0\kern-.5\wd1\fi}}
\def\centerover#1#2{\centeron{#1}{\setbox0=\hbox{#1}\setbox
1=\hbox{#2}\raise\ht0\hbox{\raise\dp1\hbox{\copy1}}}}
\def\centerunder#1#2{\centeron{#1}{\setbox0=\hbox{#1}\setbox
1=\hbox{#2}\lower\dp0\hbox{\lower\ht1\hbox{\copy1}}}}
\def\lsim{\;\centeron{\raise.35ex\hbox{$<$}}{\lower.65ex\hbox
{$\sim$}}\;}
\def\gsim{\;\centeron{\raise.35ex\hbox{$>$}}{\lower.65ex\hbox
{$\sim$}}\;}

\def\super#1{\ifmmode \hbox{\textsuper{#1}}\else\textsuper{#1}\fi}
\def\textsuper#1{\newcount\holdspacefactor\holdspacefactor=\spacefactor
$^{#1}$\spacefactor=\holdspacefactor}

\def\getcite#1,{\advance\citenumber by1
\def\getcitearg{#1}\def\lastarg{@}
\ifnum\citenumber=1
\ref{#1}\let\next=\getcite\else\ifx\getcitearg\lastarg\let\next=\relax
\else ,\ref{#1}\let\next=\getcite\fi\fi\next}

\def\pom{{\rm P\kern -0.53em\llap I\,}}
\def\spom{{\rm P\kern -0.36em\llap \small I\,}}
\def\sspom{{\rm P\kern -0.33em\llap \footnotesize I\,}}

\relax
\def\contlimit{\,{\hbox{$\longrightarrow$}\kern-1.8em\lower1ex
\hbox{${\scriptstyle (a\rightarrow0)}$}}\,}
\def\upon #1/#2 {{\textstyle{#1\over #2}}}
\relax
\renewcommand{\thefootnote}{\fnsymbol{footnote}} 

\def\mainhead#1{\setcounter{equation}{0}\addtocounter{section}{1}
  \vbox{\begin{center}\large\bf #1\end{center}}\nobreak\par}
\sectioneq
\def\subhead#1{\bigskip\vbox{\noindent\bf #1}\nobreak\par}

\def\til#1{\centeron{\hbox{$#1$}}{\lower 2ex\hbox{$\char'176$}}}
\def\tild#1{\centeron{\hbox{$\,#1$}}{\lower 2.5ex\hbox{$\char'176$}}}
\def\sumtil{\centeron{\hbox{$\displaystyle\sum$}}{\lower
-1.5ex\hbox{$\widetilde{\phantom{xx}}$}}}

\begin{document} 

\begin{titlepage} 

\rightline{\vbox{\halign{&#\hfil\cr
&\today\cr}}} 
\vspace{0.25in} 

\begin{center} 
  
{\large\bf A Pseudoscalar Resonance That
Could Resemble the Higgs}

\medskip

Alan. R. White\footnote{arw@hep.anl.gov }

\vskip 0.6cm

\centerline{Argonne National Laboratory}
\centerline{9700 South Cass, Il 60439, USA.}
\vspace{0.5cm}

\end{center}\begin{abstract} 

As described previously, in SU(5) gauge theory with massless 
left-handed $ {\bf 5 \oplus 15 \oplus 40 \oplus 45^*}$
fermions \{QUD\},
an S-Matrix is produced by infra-red anomaly dynamics that
might provide an extraordinarily economic origin and unification for the Standard Model. All particles are bound-states, with dynamically generated masses, and there is no Higgs sector. In this paper, the 
radically different dynamical role played by the top quark is elaborated. It does not reproduce Standard Model
``$~t\bar{t}~$'' events but, instead, these events are reproduced by the 
multiple vector boson decays 
of the $\eta_6$ - the remnant pseudoscalar resonance left by color sextet
electroweak symmetry breaking.
Evidence that the $\eta_6$ appears in LHC and Tevatron Z pair cross-sections is discussed. A second pseudoscalar resonance, the $\eta_3$, contains a $t\bar{t}$ pair and mixes 
with the $\eta_6$ via the pomeron. The mixing should give the $\eta_3$ 
a mass between the triplet and sextet scales. It
could also have a decay mode pattern that is more consistent with LHC and Tevatron results than the 
(mass spectrum determined) pattern of the Standard Model Higgs boson
that many hope has been discovered. Unfortunately,
at present, there is no possibility to calculate explicit cross-sections.

\end{abstract} 

\renewcommand{\thefootnote}{\arabic{footnote}} \end{titlepage} 

\mainhead{1. INTRODUCTION}

In previous papers \cite{arwdm}-\cite{arwfx}, I have outlined\footnote{I continue to work on the truly overwhelming endeavor of expanding the outline.} how a very special\cite{kw}
weak coupling massless infra-red fixed-point field theory, i.e.
\begin{center}
{ \bf QUD\footnote{QUD $\leftrightarrow$ Quantum Uno/Unification/Unitary/Underlying Dynamics}  $\equiv$} {\bf SU(5) gauge theory with   
massless left-handed 
\newline {$ {\bf 5 \oplus 15 \oplus 40 \oplus 45^*}$} fermions}
\end{center}
might provide a complete and self-contained origin and underlying unification for the Standard Model as a bound-state S-Matrix produced by massless fermion anomaly dynamics. I have arrived uniquely at QUD by asking only that high-energy cross-sections be described by the (completely) unitary Critical Pomeron\cite{cri}. It is truly remarkable that, according to the multi-regge construction that I have outlined\cite{arw10}, only Standard Model interactions are present in the QUD S-Matrix and all particles have dynamical masses - with a spectrum that may also be consistent with the Standard Model. Electroweak symmetry breaking is produced\cite{wm} by a color sextet quark sector that produces
the only new states. Moreover, solutions appear to be provided for the other core ``Beyond the Standard Model'' problems of dark matter and neutrino masses. (Indeed, the existence of small neutrino masses can be seen as a direct pointer to a very weak coupling underlying theory of just the kind that I have been led to.) 
I have described some of the experimental consequences for the LHC and for Cosmic Ray physics in \cite{arwdm}.

I have emphasized that in the QUD S-Matrix, there is no
``Higgs boson'' associated with the generation of general particle masses, although I have identified the pseudoscalar $\eta_6$, as the ``sextet higgs'' by virtue of it's relation to the sextet pion sector that generates the electroweak vector boson masses. However,
I have not previously recognized that, because of the very different nature of the QUD ``top quark'' compared to the Standard Model, there should be an additional neutral pseudoscalar resonance - the $\eta_3$. While the $\eta_3$ has a very different origin and has no direct connection with particle masses, it nevertheless may share some properties with the Standard Model Higgs boson that many would like to believe has been glimpsed at the LHC.

Initially the $\eta_3$ and $\eta_6$ are axial U(1) Goldstone bosons associated, respectively, with the triplet and sextet quark sectors. Very importantly, however, they mix via the pomeron (gluonic sector). I have previously suggested\cite{arwdm} that the $\eta_6$, which is the more massive, is responsible for what is currently interpreted as the production of $t\bar{t}$ pairs. The ``top mass'' involved is then understood 
as the characteristic mass scale of the sextet quark sector, rather than 
as an enormous triplet quark mass. A major consequence is that the $\eta_6$ should also appear in the Z pair cross-section. 

I observed in \cite{arwdm} that the four (high $p_{\perp}$) $ZZ \rightarrow llll$ events\cite{CDFzz} seen by CDF at the $t\bar{t}$ threshold could be directly due to the $\eta_6$ and, as I will discuss in Section 6, there is also much encouragement in the latest LHC data. The $\eta_6$ seems to be clearly present\cite{CMSllll} in the CMS $ZZ \rightarrow llll$ cross-section and, to a lesser extent, is visible\cite{ATLllll} in the same ATLAS  cross-section, as well as\cite{CMSllnn} the CMS ZZ $\rightarrow ll \nu\nu$ and\cite{ATLllqq} the ATLAS $ZZ \rightarrow llqq$ cross-sections. Including the Tevatron observed asymmetry\cite{CDFt} implication that a pseudoscalar resonance  could be involved in top quark production, evidence supporting the proposed role for the $\eta_6$ seems to be steadily accumulating.
In addition, the\cite{ATLllll} ATLAS $ZZ \rightarrow llll$ and\cite{ATLllll} $ZZ \rightarrow llqq$ cross-sections have significant excesses at large mass which, most likely, is the non-perturbative QUD cross-section anticipated in \cite{arwdm}.

It is fortunate that, so far, QUD appears to be able to account, experimentally, for all the intricacies of the Standard Model. In particular, the first two quark generations seem\cite{arwdm} to have properties that mesh with the known experimental situation.
It is, however, crucial for our present purpose that the top quark plays a radically different role in the QUD S-Matrix, compared to the Standard Model, even 
though (very surprisingly, perhaps) the resulting experimental phenomena may be almost indistinguishable.
It is noteworthy that while all previously discovered quarks were originally proposed as bound-state constituents, the Standard Model top quark is a very different (even bizarre!) object. It can not be identified as a constituent in any bound-state and instead exists only in the QCD lagrangian. As a result, it is given an (ambiguous at best) lagrangian mass that is two orders of magnitude larger than the typical strong interaction bound-state mass scale. 

In the QUD lagrangian all elementary quarks (and leptons\footnote{Physical leptons are bound-states that are similar to hadrons. The Standard Model lagrangian that includes quark masses might be an effective theory, for the QUD S-Matrix, in which the elementary leptons are replaced by physical bound-state leptons.}) 
have zero mass. In the S-Matrix,
effective constituent quark masses are 
acquired via the bound-states that they form. The top quark is distinguished, not 
by any large intrinsic mass, but instead by it's
electroweak coupling to a high mass exotic triplet quark sector.  Because of the absence of any chiral symmetry, the exotic sector can be expected to either not form bound-states or to only form very massive states. Consequently,
potential low mass states containing a single top quark should be destabilized. This is almost the same result as giving 
the top quark a very large intrinsic mass. The crucial difference is, however, that the neutral $\eta_3$ state that contains a 
$t\bar{t}$ pair, together with all other
triplet $q\bar{q}$ pairs, will avoid 
the mixing with the exotic states and so should have a relatively low mass - 
before the mixing with the $\eta_6$. 
 
Assuming the $\eta_6$ pulls the $\eta_3$ mass up from the triplet scale towards the sextet scale, it could have a mass comparable with the candidate Standard Model Higgs boson seen at the LHC. Also, the width will be relatively narrow if, as we will argue will be the case, it is dominated by the $t\bar{t}$ component - since the top has no direct quark decay options.
$b\bar{b}$ modes, although suppressed, would then be the most common.
There should also be (rarer) two photon and (even rarer) Z pair decay modes - via the anomaly. Leptonic decay modes, such as $\tau^+\tau^-$, would not be expected. Also if it's mass is indeed well below threshold, it will not decay via W pair production. 

It is possible, therefore, that the $\eta_3$ decay mode 
pattern could be more consistent with LHC and Tevatron results than the 
much sought after Standard Model Higgs boson. Consequently, it might actually be an alternative explanation for the current, ambiguous at best, ``Higgs discovery''.
Unfortunately, at present\footnote{Most likely, significant experimental evidence will be needed to inspire the ``community involvement'' that would rapidly 
lead to QUD becoming explicitly calculable.}, there is no possibility to calculate explicit cross-sections or decay probabilities. Nevertheless, 
if the $\eta_3$ is actually identified at the LHC, it would amount to the constituent discovery of the QUD top quark.

We can say that, because of their QCD origin, we expect the $\eta_3$ and $\eta_6$ cross-sections to be present, and to be considerably larger, away from the 
short-distance region where they might be currently seen at the LHC.
A priori, we would also expect a large variety of multiparticle decay modes - particularly
for the $\eta_3$. Regrettably, as we have lamented at length in \cite{arwdm}, the pile-up resulting from the extremely high luminosity makes it almost impossible to detect anything other than the very high $p_{\perp}$, very central, particles appearing in events. In general, and very disappointingly, there is no way to distinguish a characteristically short distance process from a kinematically fringe component of the new strong interaction provided by QUD. 

\mainhead{2. THE TRIPLET QUARK SECTOR IN QUD}

There are\cite{arwdm} two ``Standard Model'' quark generations
$$
(3,\frac{2}{3}),~ (3,-\frac{1}{3}) ~\equiv~ [3,2,\frac{1}{6}]
{\large \bf ~\in  15},  ~~~~~~ (3,\frac{2}{3}),~ (3,-\frac{1}{3}) ~\equiv~ [3,2,\frac{1}{6}] {\large \bf ~\in  40}
$$
and also an unconventional ``third generation''
$$
(3,-\frac{1}{3}) ~\equiv ~[3,1,-\frac{1}{3}] {\large \bf ~\in 5}, ~~~~
(3,\frac{2}{3}) ~\in [3,2,\frac{7}{6}] {\large \bf ~\in 45^*}
$$

Amongst the antiquarks there are two almost identical generations which, at first sight, appear to have  wrong weak interaction quantum  numbers.
$$
(3^*,-\frac{2}{3}),~(3^*,\frac{1}{3})\in [3^*,3,-\frac{2}{3}] {\large \bf ~\in 40},
~~~~(3^*,-\frac{2}{3}),~ (3^*,\frac{1}{3}) \in [3^*,3,\frac{1}{3}] {\large \bf ~\in  45^*}
$$
There is also a conventional third generation of antiquarks
$$
(3^*,-\frac{2}{3}) \equiv [3^*,1,-\frac{2}{3}] {\large \bf ~\in  40},~~~~~ (3^*,\frac{1}{3})\equiv [3^*,1,\frac{1}{3}] {\large \bf ~\in  45^*}
$$

As elaborated in \cite{arwdm} and \cite{arw10}, because anomalies are involved, the SU(2)$_L$ carried by the physical electroweak bosons is such that they
couple only to doublet SU(2) fermions and so the almost identical antiquark
pairs actually do have the right (singlet) quantum number. Moreover, since the SU(2) subgroup of the gauge symmetry is 
part of the full SU(5) reggeon interaction kernels, the triplet quantum numbers of both antiquark pairs will result in a mixing of the two pairs in a manner that could produce the mixing of the two lowest mass generations that is seen experimentally.
The physical quarks corresponding to the two antiquark pairs are discussed in \cite{arwdm}. 

The most remarkable feature of the triplet quark sector is the presence 
of a ``top quark'' in the ``unconventional third generation''. The charge 2/3 quark forms an SU(2)$_L$ doublet with a quark belonging to a set of higher charged (exotic) quarks and antiquarks 
\newline \parbox{1.2in}{$~~$ {\large \bf exotics {\Huge \{} }}
\parbox{4in}{
$$
(3,\frac{5}{3}) ~\in [3,2,\frac{7}{6}] {\large \bf ~\in 45^*}, ~~~
(3,-\frac{4}{3}) ~\equiv ~[3,1,-\frac{4}{3}] {\large \bf ~\in 45^*}
$$
$$
(3^*,-\frac{5}{3}) \in [3^*,3,-\frac{2}{3}] {\large \bf ~\in  40}, ~~(3^*,\frac{4}{3}) \in [3^*,3,\frac{1}{3}] {\large \bf ~\in  45^*}
$$}
\newline As a result, the top quark will have a physical electroweak coupling to the exotic quark sector.

\mainhead{3. THE SPECTRUM VIA SYMMETRY RESTORATION}

The QUD spectrum is obtained via the construction\cite{arwdm,arw10} of states and amplitudes in the di-triple regge region. In principle at least, we consider reggeon diagram amplitudes having the general form shown in Fig.~1.
The ``non-perturbative'' physics that produces the bound-state S-Matrix is
a consequence of anomaly chirality transitions in triangle diagram reggeon effective vertices. In the multi-regge construction these transitions are input via the removal of mass and cut-off regulators.

We begin with the gauge symmetry broken 
by both reggeon masses, that are provided by scalar field condensates, 
and a crucial $k_{\perp}$ cut-off $\lambda_{\perp}$. The fermion mass scalars are decoupled first, leaving 
chirality 
transitions that break SU(5) to the vector symmetry SU(3)$_C\otimes$U(1)$_{em}$, {\it but only in reggeon anomaly vertices.}
The subsequent successive decoupling of gauge boson scalars gives
global gauge boson reggeon symmetries $SU(2)_C \rightarrow  SU(4) \rightarrow  SU(5)$.
The last scalar to be removed is asymptotically free, allowing the limit $\lambda_{\perp} \to \infty$ to be taken before the SU(5) limit.
\begin{center}
\epsfxsize=4.8in\epsfbox{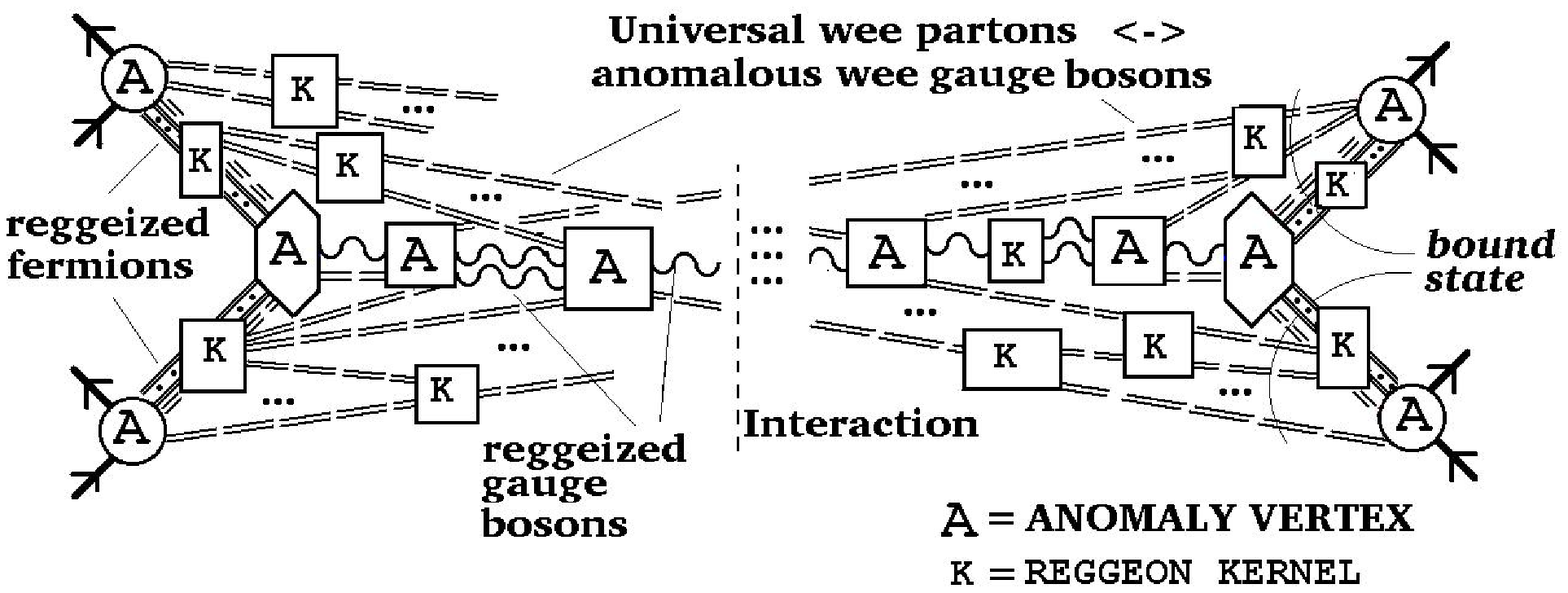}

Figure 1 A Di-Triple Regge Amplitude in QUD
\end{center}
 
After the SU(2)$_C$ symmetry restoration, the bound-states are massless Goldstone boson hadrons that are ``$q\bar{q}~$ mesons", or ``$qq$ nucleons'', or ``$\bar{q}\bar{q}~$ nucleons", with the quarks $q$ being {\bf 3's, 6's,} or {\bf 8's} under SU(3)$_C$. They are anomaly poles produced by an infra-red divergence  involving anomalous wee gluons that is a consequence of restoring the gauge symmetry while keeping a
$\lambda_{\perp}$ cut-off. The coupling of {\bf 3, 6,} and {\bf 8} pions to a scattering vector boson is illustrated in Fig.~2.
\begin{center}
\epsfxsize=5.9in \epsffile{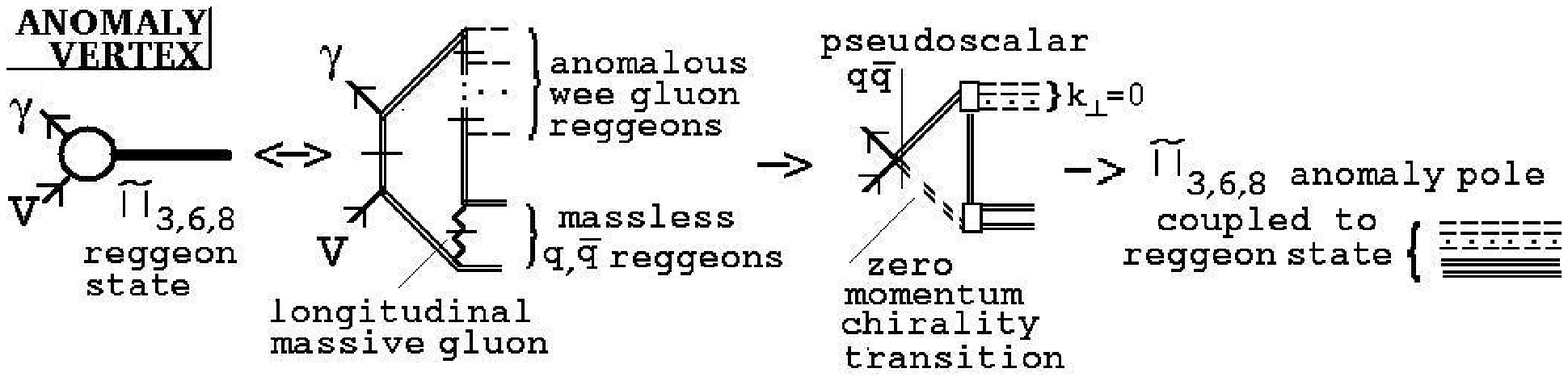}

Figure 2. An Anomaly Vertex Coupling of {\bf 3, 6} and {\bf 8} Pions With SU(2)$_C$ Restored 
\end{center}

As shown, if the bound-states are characterised by the pseudoscalar quark/antiquark state that couples directly to the scattering vector boson, then either the quark or antiquark is in a zero
momentum (``negative energy'') state. However, the bound-states are more completely characterised by the reggeon state which couples to the anomaly pole as a result of a zero momentum chirality transition. The reggeon state contains normal quark/antiquark reggeons together with an anomalous wee gluon component
that plays the role of an ``infinite momentum vacuum'' and determines the pseudoscalar nature of the anomaly pole particle associated with the reggeon state. The interactions of hadrons are obtained via the reggeon interactions of the 
corresponding reggeon states.

The {\bf 8's} have no SU(3)$_C$ anomaly, but they contain complex SU(2)$_C$ chiral doublets that produce anomaly poles when only SU(2)$_C$ is restored. After the final SU(5) symmetry restoration, the {\bf 8's} contribute only at infinite momentum.
They play a vital role in determining the physical bound-state spectrum, but they can be ignored in the present discussion. In fact, all we need to know about the limits taken after the SU(2)$_C$ limit is that the SU(2)$_L$ interaction is introduced with an electroweak scale mass and with an anomaly coupling which implies, using the quark reggeon state description of hadrons, that it couples only to quarks that are doublets under the underlying SU(2) gauge symmetry.  

\mainhead{4. MIXING OF THE {\LARGE \bf $\eta_3$} AND THE {\LARGE \bf $\eta_6$}}

With the SU(2)$_C$ symmetry restored, there will be six Standard Model
flavors of massless triplet quarks, with charges of 2/3 or -1/3, together with the 
corresponding triplet antiquarks. The chiral symmetry will result in a very large number of 
massless Goldstone boson hadrons, before all the higher order reggeon anmaly interaction mixings, that produce the particle spectrum, are taken into account. 
Because the higher charged triplet quarks and antiquarks do not have any chiral symmetry, they do not form massless chiral Goldstone bosons, even at this initial
stage. We assume, as we said above, that they either form very massive resonances or do not contribute at all to the physical spectrum. Once the electroweak interaction is introduced, the Goldstone bosons that contain a single top quark will mix with exotic sector states and so will also be either destabilized or pulled up to a very high mass scale.

A neutral $\eta_t$ meson that contains a top quark and a 
top antiquark will survive the destabilization. In the reggeon state description, the top quark and antiquark
will have opposite helicities and so, as a particle/antiparticle pair, can  form a neutral state with respect to the SU(2)$_L$ interaction and hence
avoid the coupling to the exotic sector. If we appropriately combine the $\eta_t$ with all the neutral $\eta$ mesons containing triplet quark/antiquark pairs
we obtain the $\eta_3$, in which the reggeon state will be a flavor singlet with respect to the color triplet quark sector. That is to say, initially, the $\eta_3$ is the massless axial U(1) Goldstone boson of the triplet sector. 

The ``sextet pions'' become the (dominant part of the) longitudinal component of the electroweak vector bosons and leave only the $\eta_6$ flavor singlet pseudoscalar. 
Initially, also, the $\eta_6$ is the massless Goldstone boson resulting from the sextet axial U(1) flavor symmetry. This is why, if the sextet pions are compared with the Standard Model Higgs scalars that give the vector bosons their masses, then the $\eta_6$ compares directly with the left-over scalar, i.e. the ``Higgs'',
even though the sextet sector does not duplicate the role of the Higgs sector in providing general masses.

In fact, both U(1) axial symmetries are broken by the presence of the $\lambda_{\perp}$ cut-off in the anomaly vertices. The consequent Ward identity
violation produces the infra-red divergence that results in the infra-red chirality transition of a quark/antiquark state to a reggeon state containing an anomalous wee gluon component. This is the dynamical mechanism whereby the breaking of the chiral symmetry produces a Goldstone boson.
As flavor singlets, the quark/antiquark reggeon component of both the $\eta_3$ 
and the $\eta_6$ can couple to the massive SU(2)$_C$ singlet gluon reggeon that, in combination with the anomalous wee gluon component, forms the pomeron at this stage of the symmetry restoration. The daughter of the pomeron, containing a reggeized gluon daughter\cite{ks}, will carry the appropriate spin. 
This is illustrated in Fig.~3.  
\begin{center}
\epsfxsize=2.9in \epsffile{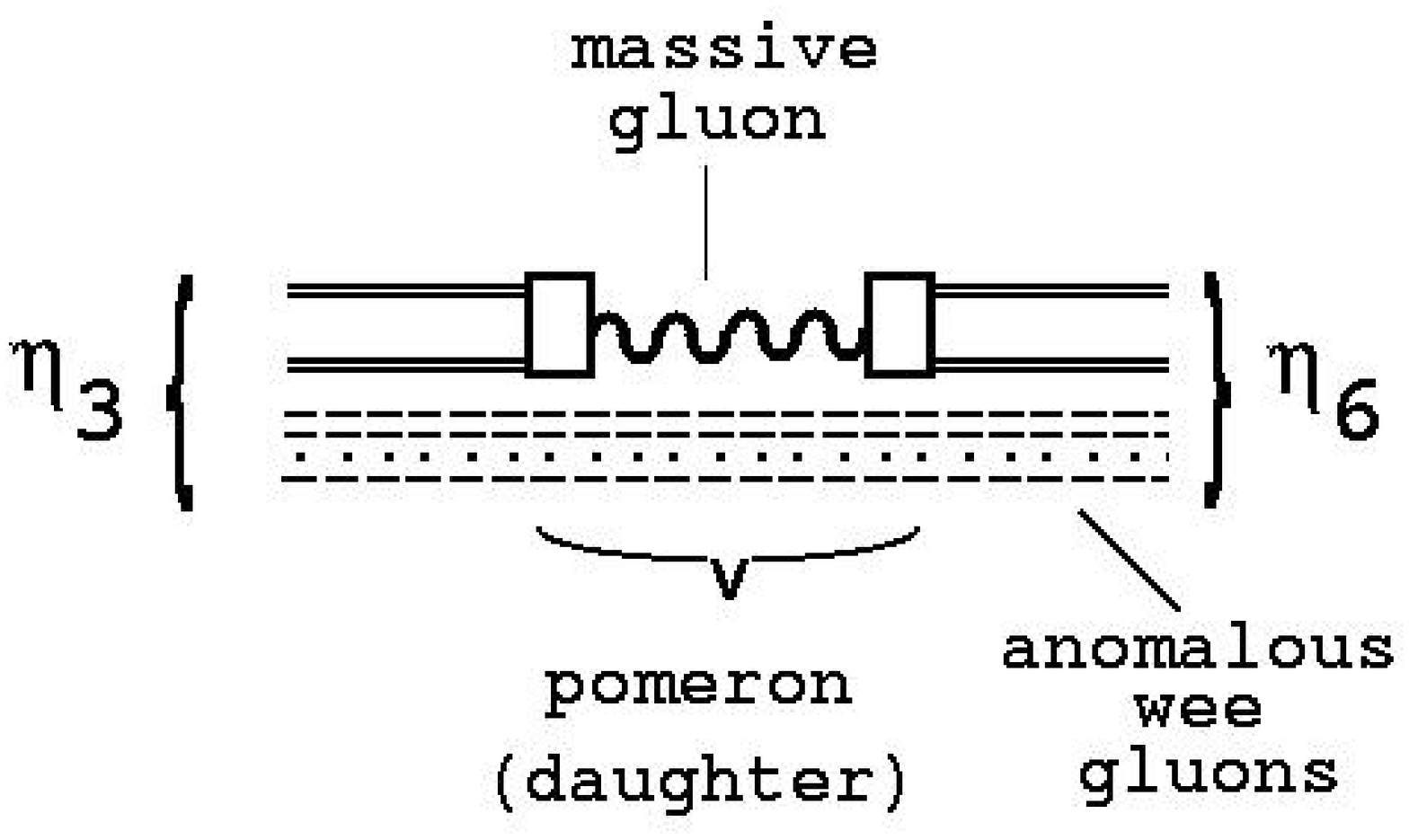}

Figure 3. Mixing of the {\large \bf $\eta_3$} and the {\large \bf $\eta_6$}
Reggeon States Via the Pomeron
\end{center}
This mixing ensures that both the triplet and the sextet U(1) symmetries
remain broken after the cut-off is removed and produces a mass for both Goldstone bosons. Because gluons can not produce anomaly pole bound states, there is no additional bound-state (glueball) produced by the pomeron daughter. As a result,
the full mixing will produce only two pseudoscalar resonances. One at the (electroweak) sextet scale - which we continue to call the $\eta_6$ - and the other, which we call the $\eta_3$, should lie somewhere between the triplet and sextet scales.

\mainhead{5. DECAYS MODES OF THE {\LARGE \bf $\eta_6$} AND THE {\LARGE \bf $\eta_3$}}

The mixing with the pomeron implies that the quark reggeon components of 
both the $\eta_3$ and the $\eta_6$ will
be produced primarily via gluon production, just as is conventionally 
assumed to be the case for the Standard Model top quark (with, however, anomalous wee gluons always in attendance). As we elaborate  below, the final states involved imply 
that the observation of a $t\bar{t}$ ``threshold'', first at the Tevatron, and now at the LHC, could
actually be the observation of the $\eta_6$.

It is commonly argued that the Standard Model top quark, once produced, has no time to interact and form a hadronic resonance before it decays. This implies that, necessarily, it is directly produced. Experimentally, however, since it is only detected by isolating potential final state events and eliminating backgrounds, there is no evidence for or against the involvement of a resonance. The appearance of a resonance at the $t\bar{t}$ threshold mass in the Z pair cross-section, as we discuss in the next Section, would be the first direct evidence.
A key consequence of the involvement of the $\eta_6$ could be that, since it is a pseudoscalar
carrying negative parity, interference with the positive parity, charge asymmetric, background produced by the initial state of the Tevatron would result in an asymmetry of the kind observed\cite{CDFt}.
 
As we have emphasized often in the past, and will return to in the 
final Section, there is much to be said, both theoretically and philosophically, in favor of substituting the dynamical mass of a sextet
quark/antiquark bound state for 
a lagrangian electroweak scale triplet quark/antiquark mass of 330 GeV. 
In particular, this avoids, altogether, the logical paradox that the mass of a colored, confined, state is a well-defined physical observable, 

Since we are discussing strong interation physics, we
expect that the decay modes of the $\eta_6$
have to be described non-perturbatively. Moreover, because of the distinctive 
nature of the top quark, the Standard Model perturbative description of 
decays is obviously not applicable. Fortunately, many non-perturbative experimental features are similar to the perturbative picture.

Assuming the $\eta_6$ is primarily a sextet quark bound-state, we start by exploiting the parallel between the \{$\pi^{\pm}_6,\pi^0_6,\eta_6$\} sextet
states, corresponding to \{$W^{\pm},Z^0,\eta_6$\},
and the familiar \{$\pi^{\pm},\pi^0,\eta$\} triplet quark states.
Since higher-scale strong-interaction physics is involved, we expect the width to
be large (in contrast to the $t\bar{t}$ dominated $\eta_3$ that we discuss next). If we take 
$m_{\eta_6} \sim 2 m_{top} \sim$ 330 GeV, then the relative couplings and masses of
the vector mesons, and the photon, imply that the 
primary non-perturbative decay mode should be (in parallel with 
$\eta~\to~ \pi^+~\pi^-~\pi^0$) 
$$
\eta_6~~\to~~ W^+~W^-~Z^0 ~~~\to~~W^+~W^-~b\bar{b}
\auto\label{dk1}
$$
which, if $Z^0 \to b\bar{b}$ as shown, would give (perhaps remarkably) exactly the same final state as the dominant perturbative Standard model $t\bar{t}$ decays. 
The next most significant mode 
$$
\eta_6~~\to~~ Z^0~Z^0~Z^0 
\auto\label{dk2}
$$
(in parallel with $\eta~\to~ \pi^0~\pi^0~\pi^0$) 
should have a smaller branching ratio because of the larger $Z^0$ mass. 
In fact, when the $Z^0$'s decay hadronically, as they
do most of the time,  (\ref{dk2}) would again  
be indistinguishable from (\ref{dk1}). (This would also be the case in most 
leptonic decay modes.)

If the parallel between the \{$\pi^{\pm}_6,\pi^0_6,\eta_6$\} sextet
states components of \{$W^{\pm},Z^0,\eta_6$\} and the
 \{$\pi^{\pm},\pi^0,\eta$\} triplet quark states were complete then 
the decay modes 
$$
\eta_6~~\to~~ W^+~W^-, ~~~~\eta_6~~\to~~ Z^0~Z^0
\auto\label{dk21}
$$
would parallel $ \eta \to \pi^+ \pi^-$ and $ \eta \to \pi^0 \pi^0$ and so would 
be forbidden, as non-perturbative QCD decays, by parity. In fact,
because the final states are vectors and not pseudoscalars this argument does not hold. Nevertheless, the decays are likely to be relatively suppressed since they can not go so directly through sextet meson interactions. Most importantly, however,
the $Z^0Z^0$ decay can also proceed electromagnetically via the anomaly (in
parallel with  $\eta \to \gamma \gamma$). 

Because the $\eta_6$ mass is so large, radiative
decay modes, such as 
$$
\eta_6~~ \to~~ W^+~W^-~\gamma~, ~~~Z^0 ~Z^0~\gamma ~, ~~~ Z^0 ~\gamma~\gamma~, 
~~~ \gamma~\gamma~,
\auto\label{dk3}
$$
that require an electromagnetic coupling, will have much smaller branching ratios,
although $\eta_6 \to \gamma\gamma$ can also go, of course, via the anomaly.

Initially, after only the SU(2)$_C$ gauge symmetry is restored, all flavor $q\bar{q}$ pairs contribute equally to the $\eta_3$, in order to form the flavor neutral state that couples to the pomeron. As the full gauge symmetry is restored and all the physical states and interactions mix, the relative flavor components of the $\eta_3$ will change as part of the process whereby constituent quark masses
are developed. Assuming that the physical bound states involving the five flavors,
apart from the top quark, develop the mass spectrum seen experimentally then only the $t\bar{t}$ component can be significantly involved in the mixing with the 
$\eta_6$. This is consistent with the dramatic effect that the SU(2)$_L$ 
electroweak interaction is expected to have on the $t$ and $\bar{t}$ bound states.
We assume, therefore, that the mixing with the pomeron creates an $\eta_6$ 
that has a primary sextet component with only a small $t\bar{t}$ component. Conversely,
the $\eta_3$ will have a primary $t\bar{t}$ component, together with a small sextet 
component and a component containing the other triplet $q\bar{q}$ pairs that will primarily be $b\bar{b}$. 

With these assumptions, if the $\eta_3$ has a mass that is well below the $\eta_6$ mass, it will decay as described in the Introduction. It will not be 
able to decay via the sextet component, since it will be below the mass of all sextet states. It will surely have anomaly decays (via both triplet and sextet
quarks) into two photons and also into a
(relatively suppressed) off-shell ZZ pair. We would expect, however, that it will predominantly decay via the small $b\bar{b}$ 
component - with a  relatively small decay width. Similar reasoning would suggest that it also has a relatively small production cross-section.

\mainhead{6. THE {\LARGE \bf $\eta_6$} IN ZZ CROSS-SECTIONS}

The possibility that the experimental events assumed to be $t\bar{t}$ production are actually due to $\eta_6$ resonance production of multiple electroweak vector
bosons is a dramatic consequence of the underlying presence of QUD, when compared with the Standard Model. Analyzing these events in detail to distinguish this possibility from the conventonal picture could be
quite challenging. A more precise understanding of the QUD sextet sector is,
presumably, a necessary prerequisite. Whether there is a resonance present in Z pair cross-sections at the $t\bar{t}$ threshold mass and, in addition, whether there is a related excess in this cross-section at high mass, seems more straightforward to determine. In fact, the most recent LHC results are very encouraging, although somehat disturbing differences appear when we compare CMS and ATLAS cross-sections. The differences could, perhaps, highlight the difficulty of extracting these cross-sections in the ultra-high luminosity environemt.

We begin with the Tevatron results that were published last year and discussed in \cite{arwdm}. We show, in Fig.~4, the CDF\cite{CDFzz} $ZZ \to llll$ mass spectrum,  
together with the comparable D0 spectrum.
The four closely clustered CDF events are  
(within experimental accuracy) right at the $t\bar{t}$ threshold and clearly suggest the presence of a resonance. The D0 spectrum is consistent if the resonance is actually quite broad. In these plots, and all the subsequent plots, I have indicated both the potential location of the $\eta_6$, and the width that is suggested by the consistent combination of all plots.
\begin{center}
\epsfxsize=2.6in\epsffile{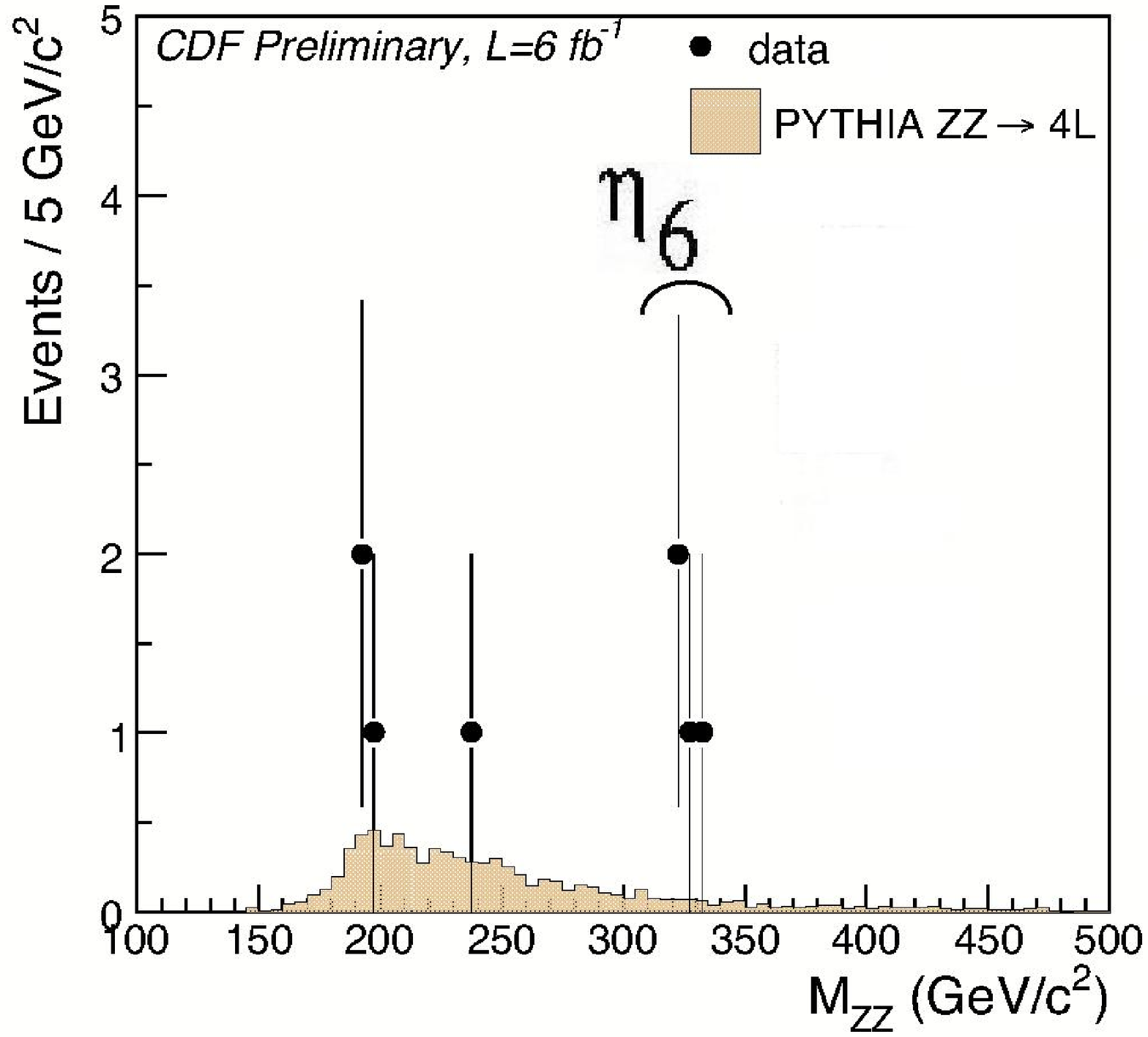}
\hspace{0.5in}
\epsfxsize=2.3in\epsffile{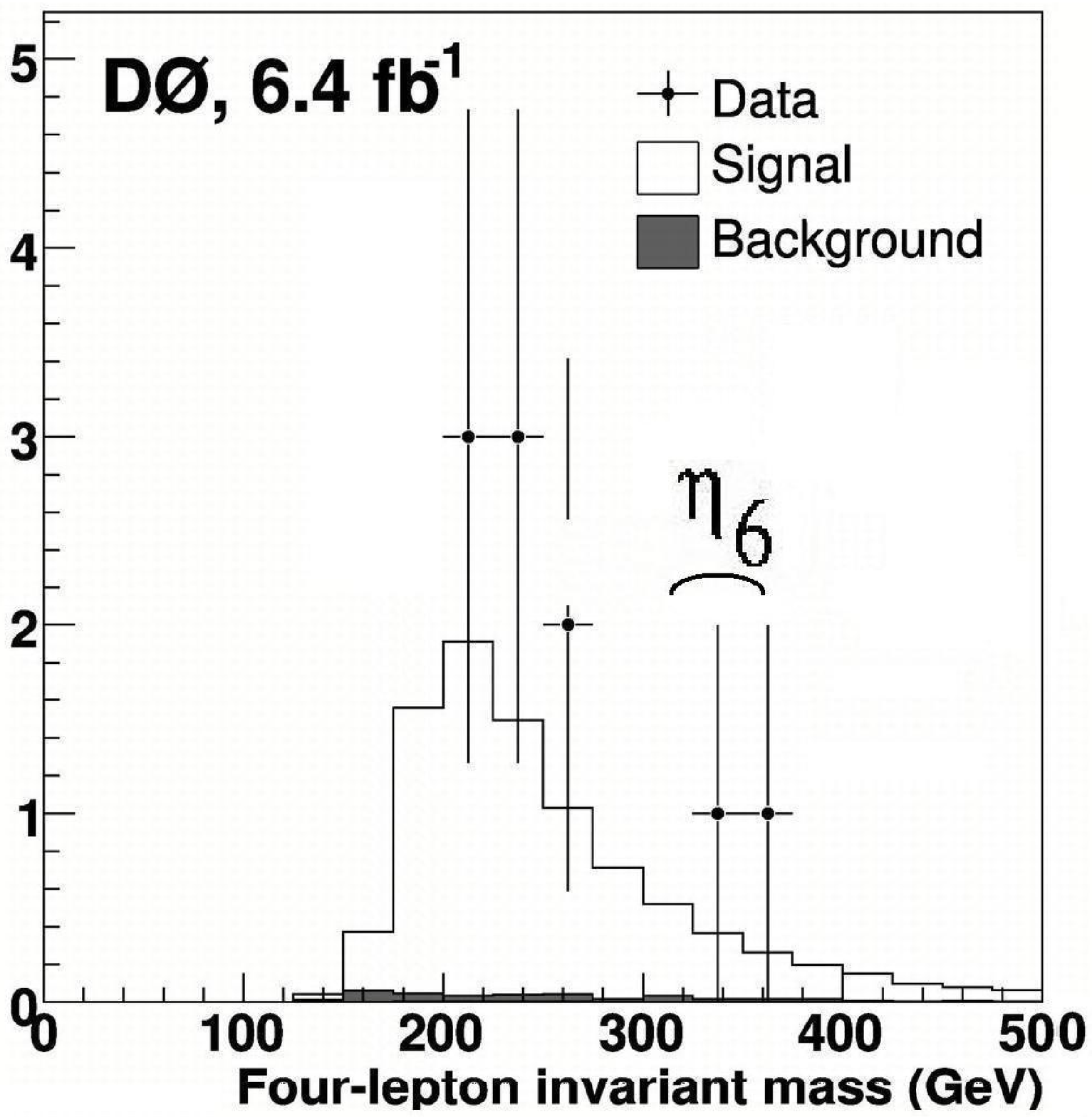}

(a) \hspace{2.8in} (b)

Figure 4. ZZ $\to ~llll$  (a) CDF mass spectrum (b) D0 mass spectrum
\end{center}

CMS results are shown in Fig.~5. The $\eta_6$ seems to be clearly present in 
the\cite{CMSllll}
$ZZ \to llll$ cross-section. From Fig.~5(a) we see that there is a total of 
nine events in the peak and from Figs.~5(b) and 5(c) we see that it is estimated to be a $2\sigma$ effect. 
Also, from Figs.~5(a) and 5(c), the peak seems better established than any other, in terms of the number of events involved.
The CMS data for\cite{CMSllll} $ZZ \to ll\nu\nu$ is shown in Fig.~5(d). There is, again, a signicant number of events in the $\eta_6$ peak, although the background is larger. It is notable that some of the $ZZ \to llll$ events have high 
$p_{\perp} (ZZ)$, as is also the case for the CDF events.
\begin{center}
\epsfxsize=3in\epsffile{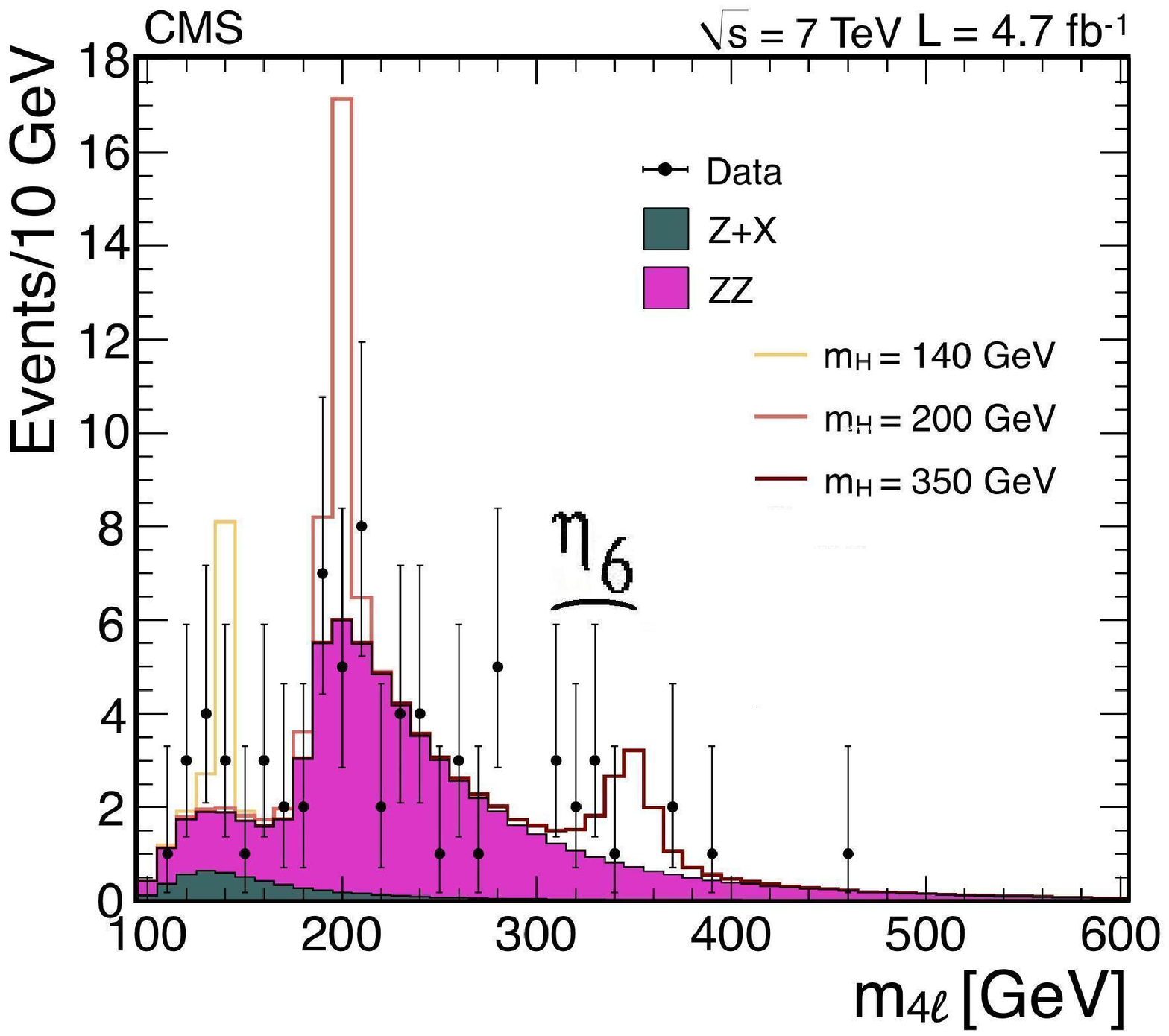}
\hspace{0.3in}
\epsfxsize=2.5in\epsffile{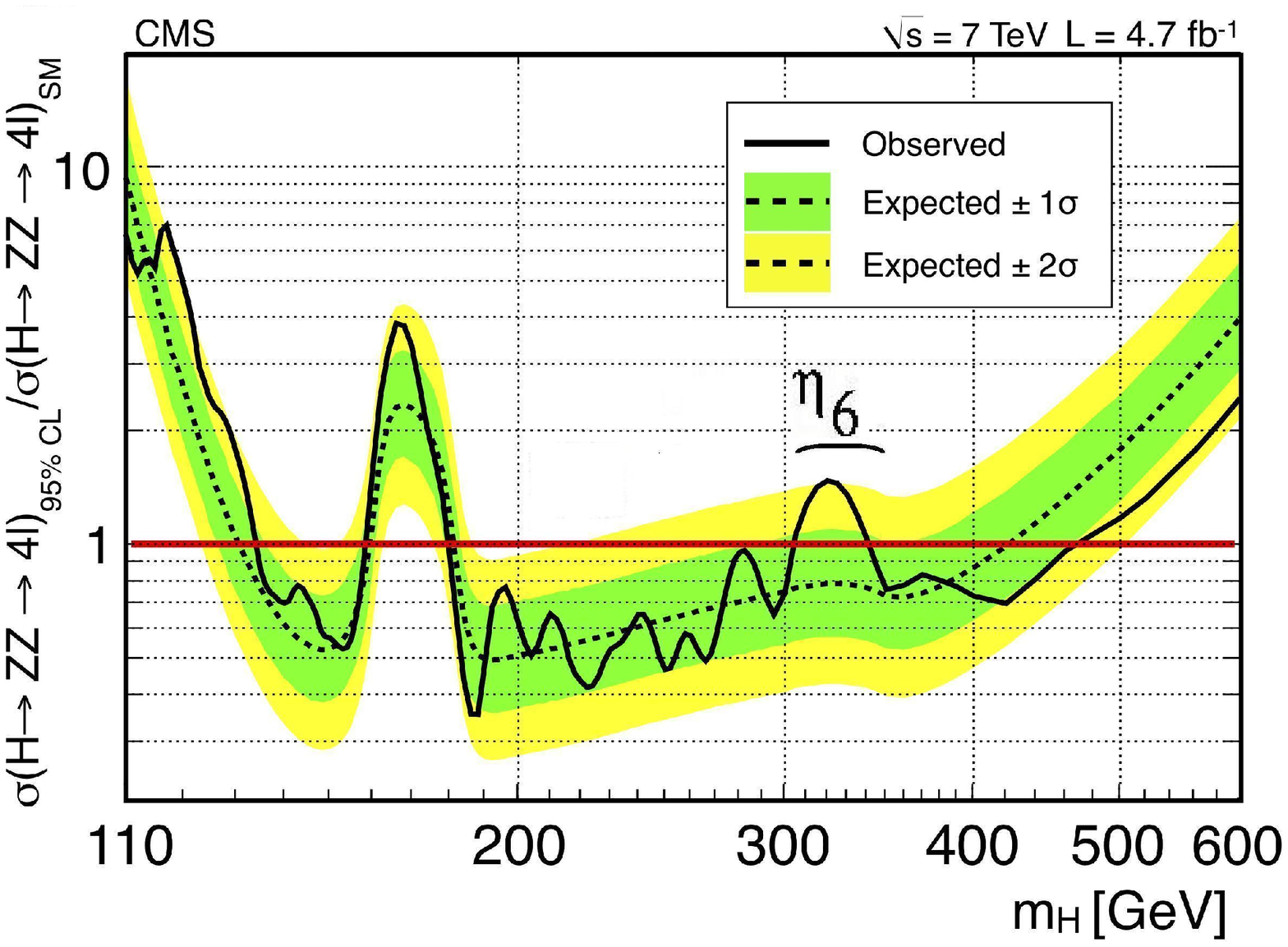}

(a) \hspace{2.8in} (b)

\epsfxsize=2.9in\epsffile{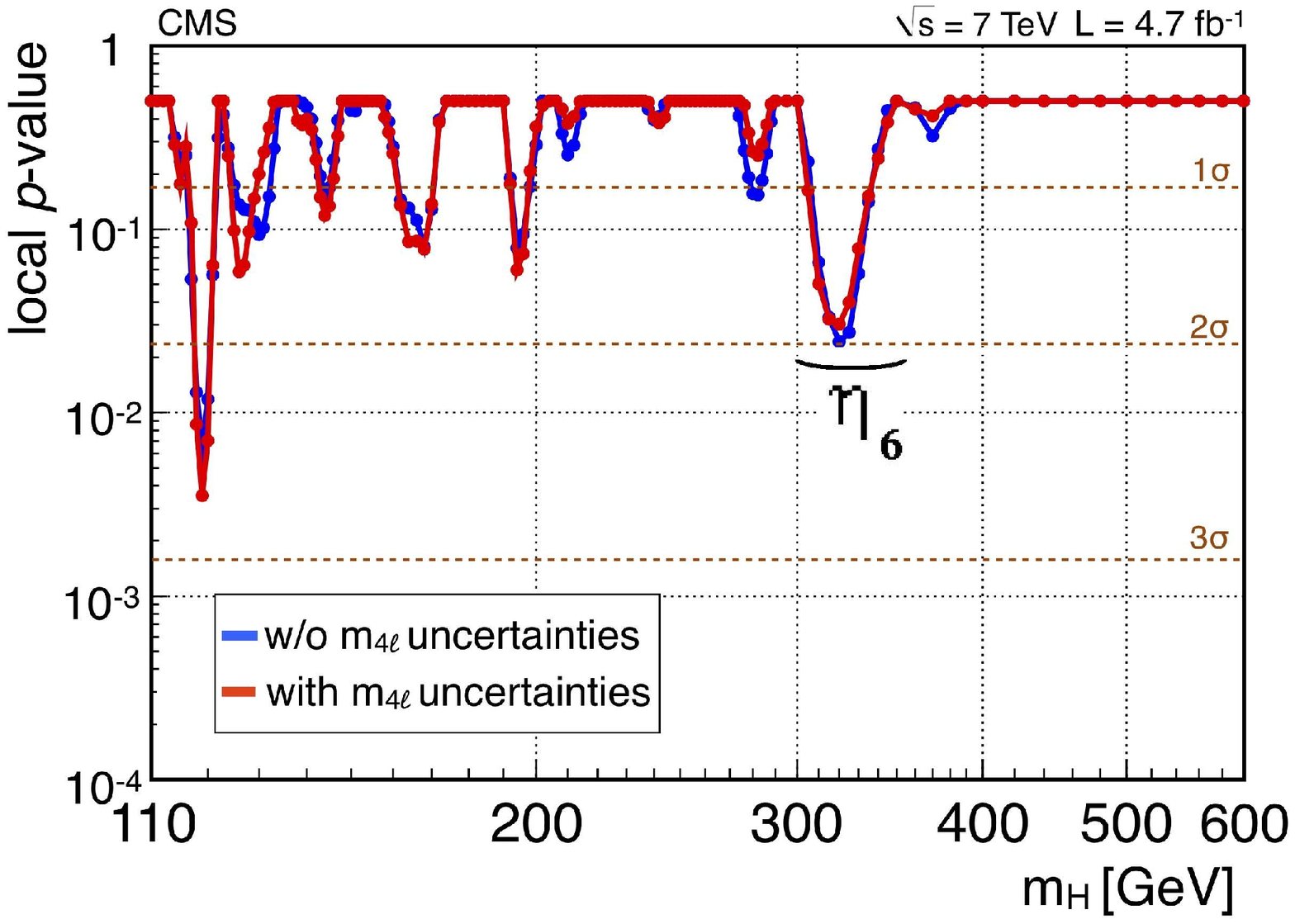}
\hspace{0.3in}
\epsfxsize=2.5in\epsffile{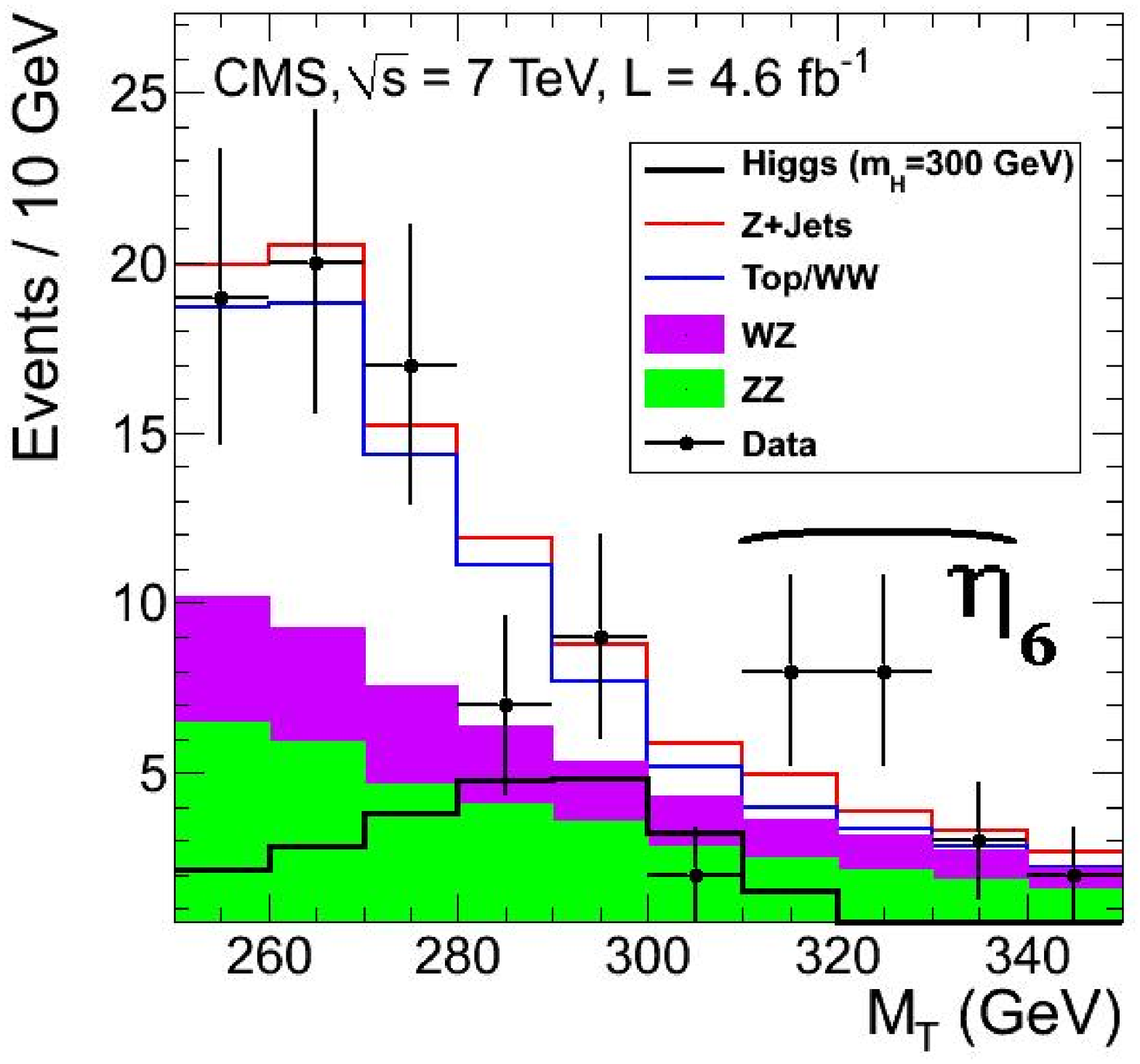}

(c) \hspace{2.8in} (d)

Figure 5. CMS (a) ZZ $\to ~llll$ Mass Spectrum ~(b) ZZ $\to ~llll$ Standard Model Comparison (c) ZZ $\to ~llll$ p-Value (d) ZZ $\to ~ll\nu\nu$ Mass Spectrum
\end{center}

ATLAS results for\cite{ATLllll} $ZZ \to llll$ and\cite{ATLllqq} $ZZ \to llqq$ are shown in Fig.~6. A somewhat different picture emerges. In both processes the $\eta_6$ appears to be weaker than in the CMS
data  but (almost by way of compensation) there are instead significant large mass excesses. The excess at 500 GeV in the  $ZZ \to llll$  cross-section is again estimated to be a $2\sigma$ effect.
\begin{center}
\epsfxsize=2.7in\epsffile{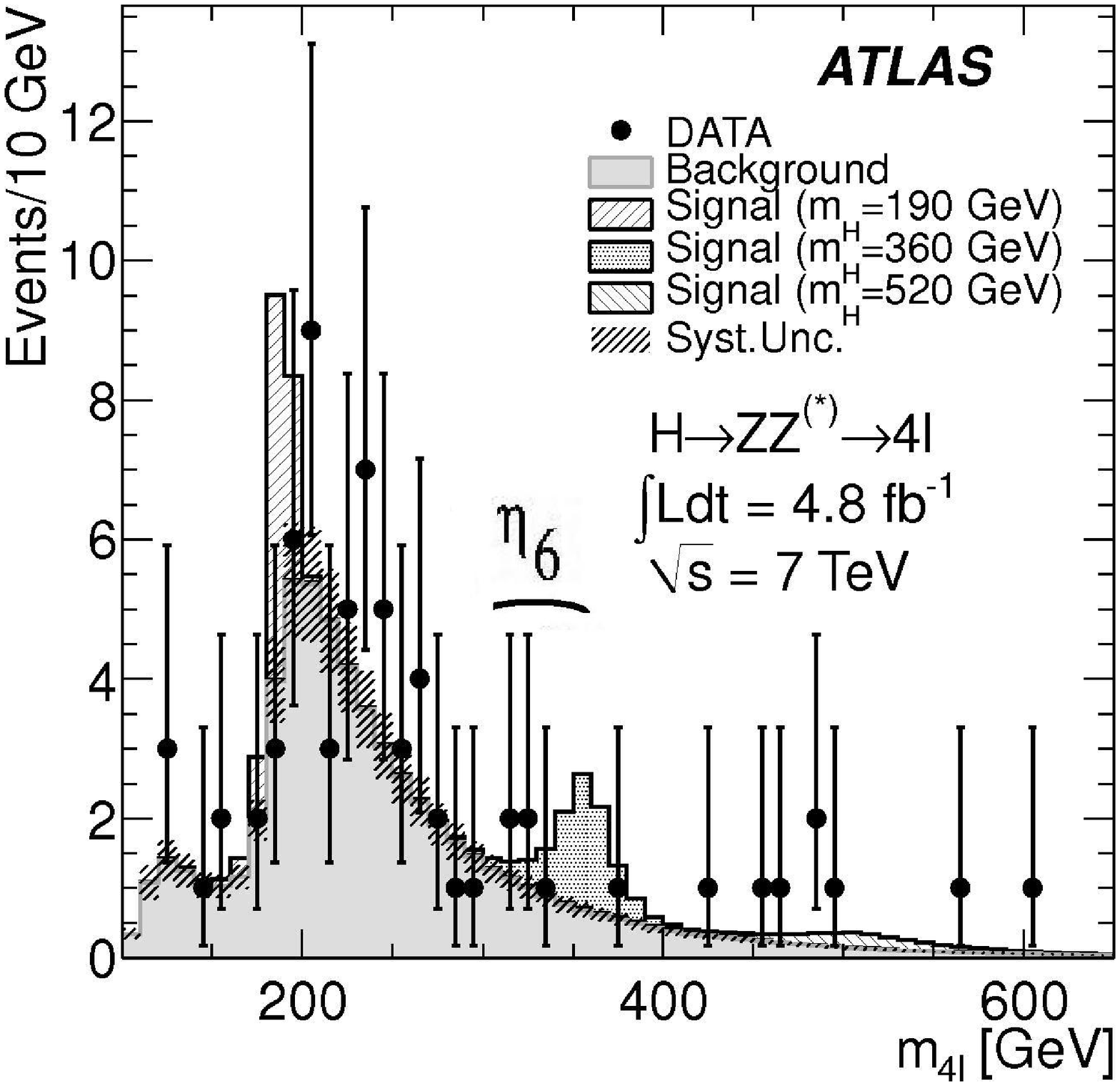}
\hspace{0.3in}
\epsfxsize=2.6in\epsffile{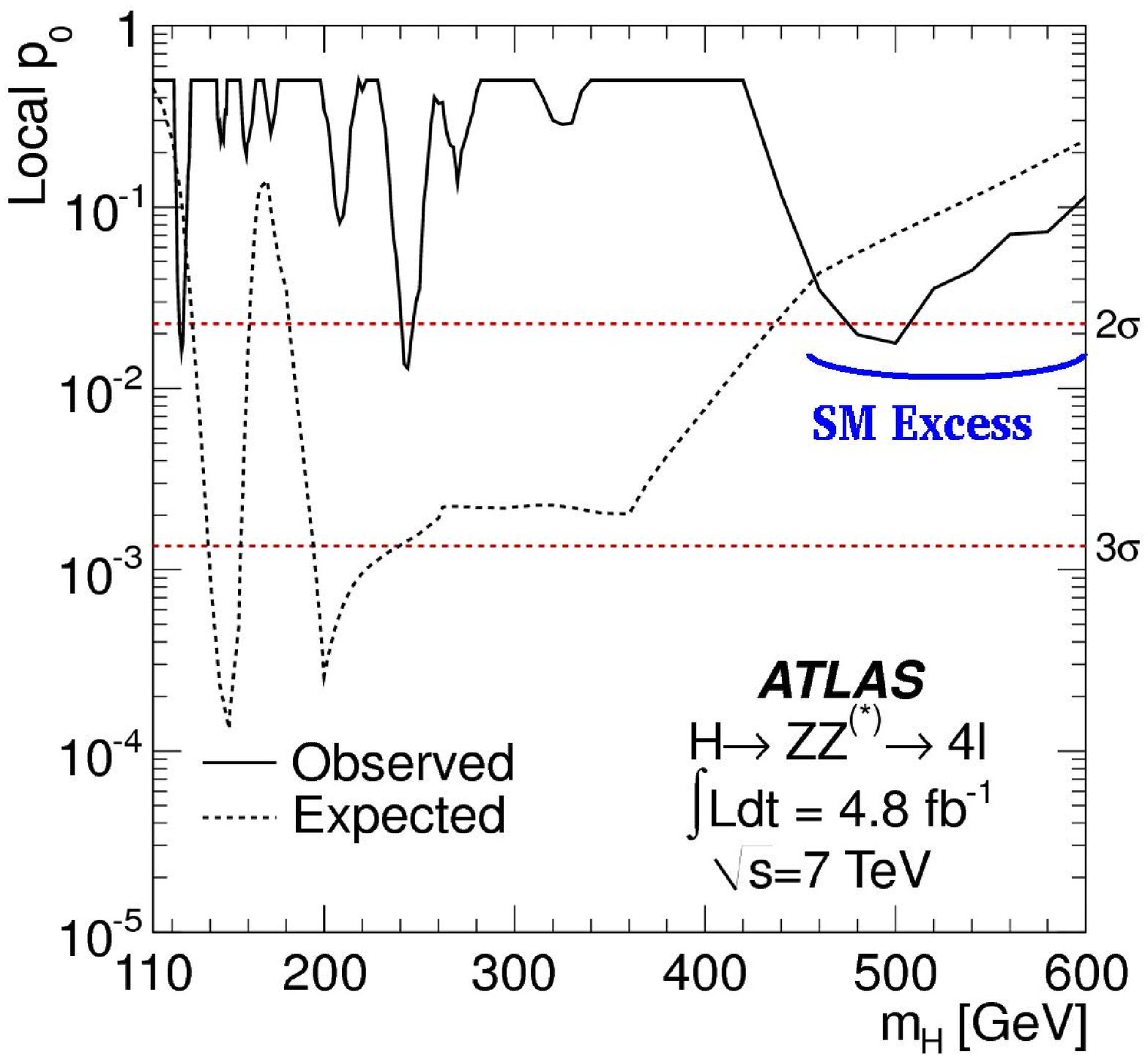}

(a) \hspace{2.5in} (b)

\epsfxsize=2.7in\epsffile{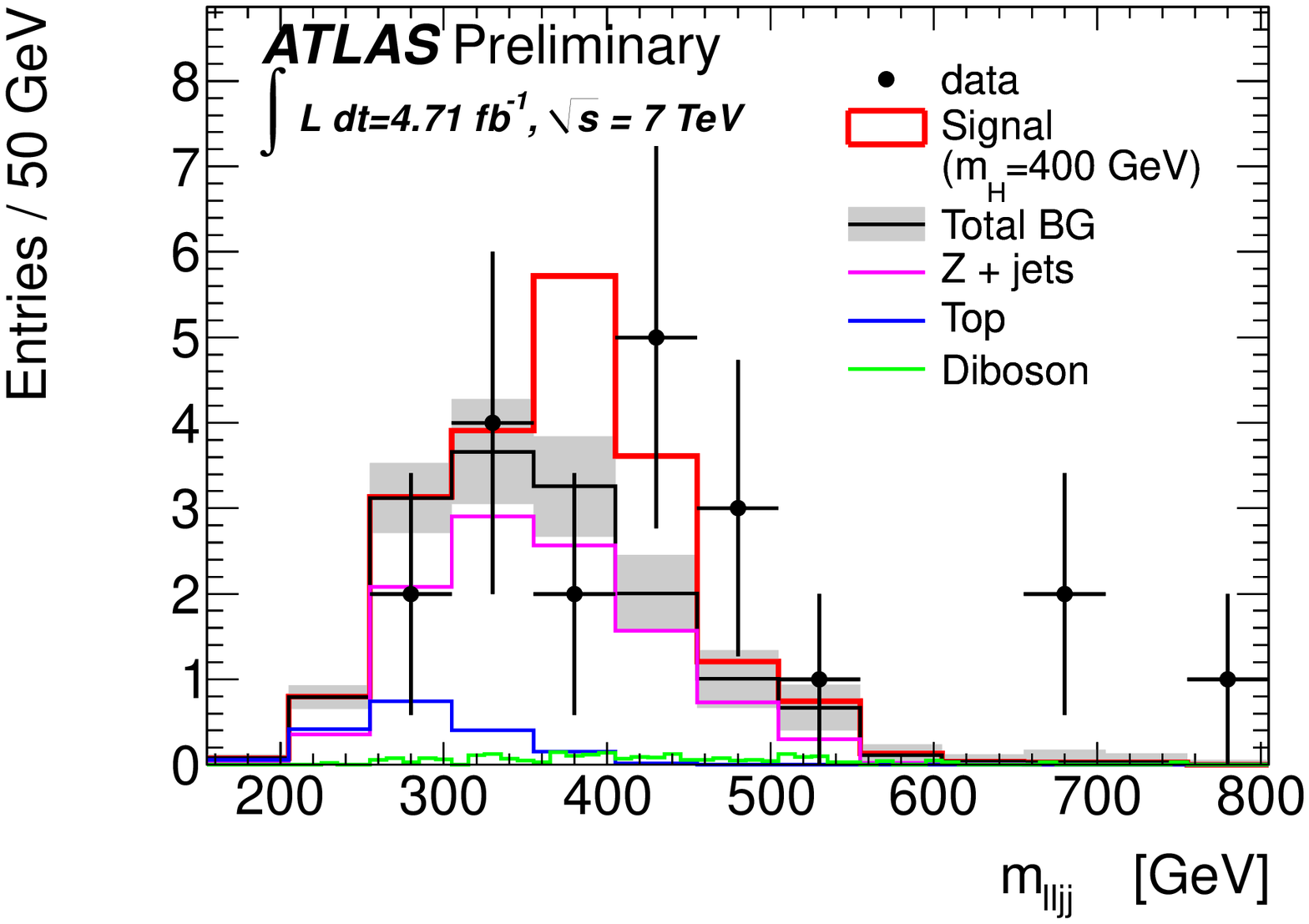}
\hspace{0.3in}
\epsfxsize=2.9in\epsffile{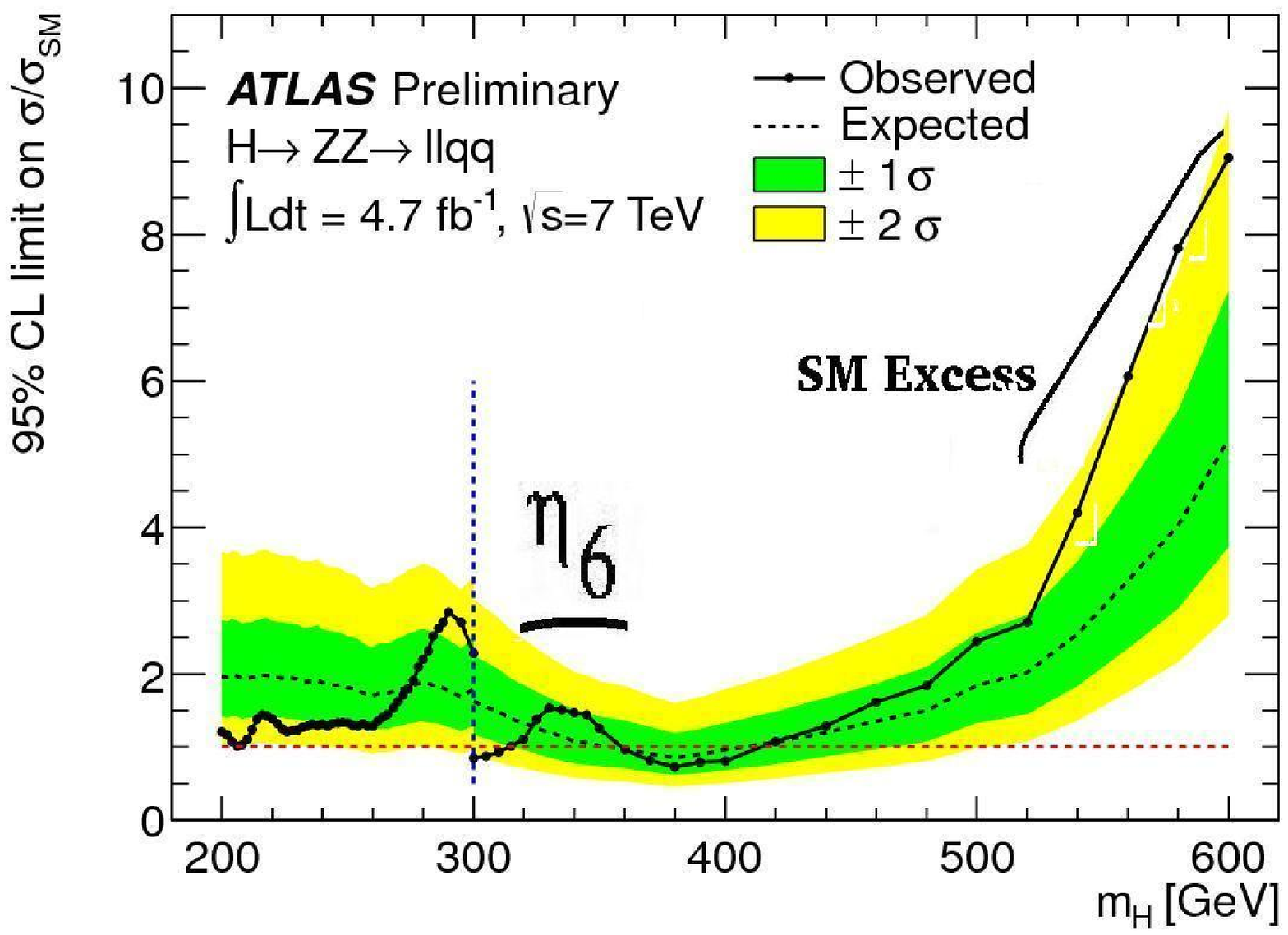}

 (c) \hspace{2.5in} (d)

Figure 6. ATLAS (a) ZZ $\to ~llll$  Mass Spectrum ~(b) ZZ $\to ~llll$ p-Value 
\newline (c) ZZ $\to ~llqq$ mass spectrum
(d) ZZ $\to ~llqq$ Standard Model Comparison
\end{center}

Apparently, the two LHC experiments provide separate, distinct, evidence for the two phenomena we would like to see. In fact, there is a related difference between the two sets of data that is somewhat disturbing. Between 300 and 400 Gev CMS has 12 $ZZ \to llll$ events, with 9 lying within the 
$\eta_6$ peak. Above 400 GeV, CMS has only one $ZZ \to llll$ event and hence have no high mass
excess. In contrast, between 300 and 400 GeV ATLAS has only 6 $ZZ \to llll$ events, with 5 lying within the $\eta_6$ peak. Above 400 GeV they have 8 $ZZ \to llll$ events and hence they have a significant high mass excess. In the 300-400 GeV mass interval, the significance of the $\eta_6$ events relative the the total number is, essentially, the same in the two experiments. However, the difference in the overall mass distribution leads to a different conclusion on the excess relative to the Standard Model. 

The quantitative difference in the high mass distributions of the two experiments 
is signigicant enough that it seems hard to ascribe it to the fickleness of low statistics data. It 
seems that it must be due to a systematic difference between either
the detection abilities of the two detectors, or the event analysis that selects the events, or both. If this is the case, then these are disturbing possibilities.

\mainhead{7. EPILOGUE - THE MISBEGOTTEN TOP QUARK ?}

If QUD does underly the Standard Model then, necessarily, the Higgs boson will not be discovered at the LHC (although the $\eta_3$ could be) and there will continue to be no sign of any superstructure associated with supersymmetry, extra dimensions, technicolor, etc. This will surely intensify the, already existing, introspection asking where the theory paradigm underlying the general thrust and direction of particle physics has failed.
In this short comment I will suggest that the conventional incorporation of the top quark into the Standard Model may have been a major contributor 
to the misdirection.

The formulaton  of the Standard Model was based on properties of the strong and  electroweak interactions that were initially discovered at relatively low energy 
(relatively long distance).
Nevertheless, as I elaborated in \cite{arw08}, the discovery of asymptotic freedom, 
together with the proven renormalizability of the electroweak interaction,
convinced most physicists that short-distance physics is ``fundamental'' for the formulation of potential physical theories while 
long-distance physics, particularly unitarity, is little more than an accounting problem that will
take care of itself. This led, in turn, to the general belief that ``new physics'' would be found via the discovery of new short-distance interactions belonging to an extended theory. Even though, as I emphasized, there was no evidence and no historical precedent to support this belief.
 
The addition of the conventional top quark to the Standard Model had a number of consequences. First, it said that a field should be added to the QCD lagrangian that has a physical effect only at very short distance 
(cancelation of the electroweak anomaly being a major justification)
and so can only be detected via interaction events. More importantly, perhaps, it implied that conventional QCD is straightforwardly operative beyond the electroweak scale and, given that the 
large top quark mass must be related to other phenomena, it confirmed
that new physics should be found by studying interaction events at high scales. This opened the door to 
the much-desired hypothesis that extra fields, for the supersymmetrization of QCD say, would appear at the electroweak scale, and higher, that could only be discovered via short-distance (large $p_{\perp}$) events. In effect, the high mass Standard Model top quark strongly encouraged existing
theoretical prejudices.

I argued in \cite{arw08} that, instead, the way forward should involve addressing 
the fundamental long-distance QCD problems 
of confinement, the origin of the parton model, and the absence of 
glueballs. I said that ignoring these problems has led, in part, to a wide variety of elaborate and unconstrained extended theories, none of which has made any contact with experiment.
(This statement was made before the LHC began operation but, of course, it becomes 
much stronger with the LHC results included.)
In contrast, the pursuit of long-distance (small $p_{\perp}$) high-energy unitarity 
via the Critical Pomeron leads {\it uniquely} to the resolution of the fundamental QCD problems, a more sophisticated electroweak anomaly cancelation, and finally,
the possible origin of the Standard Model in QUD. 

If all the conventional wisdom and prejudice, and the associated paradigm, could somehow have been suspended, other possibilities for the significance of the ``top quark events'' (and for anomaly cancelation) might, conceivably, have been considered. Nevertheless, it would have been hard to imagine that these events do not point to any of the short-distance superstructures provided by theoretical invention, but instead are a signal of a new strong interaction scale for the production of electroweak vector bosons. If, indeed, QUD is present as an underlying theory, it will ultimately lead to 
a scientific revolution in which higher-energy ``long-distance physics'' overwhelms the currently prevalent  ``short-distance'' dogma!
The physics of QUD is actually relatively simple, but the paradigm shift away from conventional field theory expectations is dramatic\cite{arwdm}-\cite{arwfx} \{{\it see the quotes below}\}.
The most significant being that the QUD S-Matrix is a necessary and sufficient description of particle physics that is isolated and self-contained, as I have suggested may be the case. Even if right, the philisophical change involved  will inevitably 
take a  very long time to be accepted\footnote{It would 
surely fuel the Weinberg/Kuhn controversy about the significance of 
paradigm change - http://www.physics.utah.edu/~detar/phys4910/readings/fundamentals/weinberg.html}.

\subhead{The Paradigm Shift - Quotes} 

From \cite{arwfx} - {\it ``the origin of the Standard Model as a bound-state S-Matrix embedded .. in an almost conformal massless field theory .. is a 
radical proposition which the LHC will determine to be either crazy heresy or singularly original insight. ... The discovery of QUD at the LHC would have a revolutionary effect on the field !!''}

From \cite{arw08} - {\it ``The S-Matrix anomaly physics is conceptually and philosophically radical
with respect to the current theory paradigm. ... QUD 
could provide a welcome way out of
the current ``Crisis in Fundamental Physics'' with, potentially, it's
novel physical applicability resolving a variety of Standard
Model problems.''}

\end{document}